\documentclass[seceq]{ptptex}
\usepackage{wrapft}
\usepackage{graphicx}

\title{%
Mapping Model of Chaotic Phase Synchronization
}
\author{
Hirokazu \textsc{Fujisaka}$^{1,}$\footnote{Email: fujisaka@i.kyoto-u.ac.jp}, 
Satoki \textsc{Uchiyama}$^{2,}$\footnote{Email: uchiyama@amath.hiroshima-u.ac.jp}, and
Takehiko \textsc{Horita}$^{3,}$\footnote{
Present address: Department of Mathematical Sciences, 
Osaka Prefecture University, Sakai,\\
$\qquad \;\;$ Osaka 599-8531, Japan, 
Email: horita@ms.osakafu-u.ac.jp}
}
\inst{
$^1$
Department of Applied Analysis and Complex Dynamical Systems,\\
Graduate School of Informatics,\\
Kyoto University, Kyoto 606-8501, Japan\\
$^2$
Department of Artificial Complex Systems Engineering,\\
Hiroshima University, Higashi-Hiroshima 739-8527, Japan \\
$^3$
Department of Mathematical Informatics,\\
Graduate School of Information Science and Technology,\\
The University of Tokyo, Bunkyo-ku, Tokyo 113-8656, Japan
}

\abst{
A coupled map model for the chaotic phase synchronization and its 
desynchronization phenomenon is proposed.  The model is constructed by 
integrating the coupled kicked oscillator system, kicking strength depending 
on the complex state variables.  It 
is shown that the proposed model clearly exhibits the chaotic phase 
synchronization phenomenon.  Furthermore, we numerically prove that in the 
region where the phase synchronization is weakly broken, the anomalous scaling 
of the phase difference rotation number is observed.  This proves that the 
present model belongs to the same universality class found by Pikovsky 
{\it et al.}.  Furthermore, the phase diffusion coefficient in the de-synchronization 
state is analyzed.
}

\begin{document}
\maketitle
\section{Introduction}
Coupled chaos systems are not only theoretically interesting but also 
important from the viewpoint of the application to engineering and 
biology~\cite{PRK01,MMP02,AAN02}.  Dynamical 
behaviors observed in coupled oscillator systems depend on the number of 
chaotic oscillators as well as the coupling form.  As the number of 
oscillators is increased, the variety 
and the complexity of dynamical behaviors increase.  One of the eminent 
characteristics of the 
coupled chaos systems is the synchronization and its breakdown.  For a coupled 
system consisted of identical chaotic oscillators, it was shown that the 
oscillators can completely synchronize under certain conditions~\cite{FY83}.  
This phenomenon is called either the chaotic complete synchronization or 
the chaotic Huygens phenomenon~\cite{FY83}.  
On the other hand, even when two oscillators have a mismatch in characteristic, 
they can show the so called chaotic phase synchronization for a certain region of 
coupling strength~\cite{RPK96}.  

It is well known that the coupled map system played a significant role in the 
study of complete synchronization~\cite{YF83,K90}.  Particularly, coupled 
one-dimensional map system is a typical model in discussing the chaos 
synchronization-desynchronization 
phenomenon.  In contrast to the complete synchronization, coupled mapping model 
for chaotic phase synchronization-desynchronization phenomenon is not studied so 
much because of the lack of suitable mapping model for coupled 
phase synchronization.  
The aims of the present paper are to propose a coupled mapping model 
for chaotic phase synchronization and then to examine the universality 
of the breakdown of synchronization so far mainly studied for the 
coupled R\"{o}ssler system.

The present paper is constructed as follows.  In Sec.~2, we propose a general, 
linearly coupled mapping system for coupled chaotic oscillators which show 
either periodic or chaotic characteristics.  The model is constructed by 
integrating the equations 
of motion for oscillator variables with a 
state dependent kicking term.  In Sec.~3, using a two oscillators system, 
we discuss general characteristics of the system, and then 
study the phase difference statistics associated with the breakdown of 
the chaotic phase synchronization.  
We give summary and concluding remarks in Sec.~4.
\section{Mapping model of coupled oscillatory chaotic systems}
First, consider the linear equations of motion
\begin{eqnarray}
\dot{\psi }(t)&=&p,\\
\tau \dot{p}(t)&=&i\omega \psi -p,
\end{eqnarray}
where $\psi $ and $p$ are complex variables, $\tau $ is a positive number 
and $\omega $ is a real number.  In the limit $\tau \rightarrow 0$, 
the above equations 
reduce to $\dot{\psi }=i\omega \psi $, the equation of motion for a 
harmonic oscillator.  So, $\omega $ has the meaning of the 
eigenfrequency of the oscillator, and the above set of equations of 
motion turns out to describe a harmonic oscillator slightly modulated by 
the introduction 
of the ``inertia term'' $\tau \dot{p}$.  The above equations of 
motion can be solved as $\psi (t)\sim e^{\lambda t},\ p(t)\sim e^{\lambda t}$, 
with the characteristic exponents  
$\lambda =\tau^{-1}(-1\pm \sqrt{1+4i\tau \omega })/2$.  For small 
$\tau $, they are written as 
$
\lambda = i\omega +\tau \omega^2,\ 
-\tau^{-1}-i\omega .
$
One should note that in the presence of the inertia term, the above harmonic 
oscillation is unstable.
\par
We consider the periodically-kicked harmonic oscillator system
\begin{eqnarray}
\dot{\psi }(t)&=&p,\\
\tau \dot{p}(t)&=&i\omega \psi -p+F_a(\psi ,\psi^*)\cdot 
\sum_{n=-\infty }^{+\infty }\delta (t-t_n),
\end{eqnarray}
$(t_n=nT)$, where $T$ is the period of kicking and is taken to be unity 
without loss of generality.  $F_a(\psi ,\psi^*)$ is the state-dependent 
complex amplitude 
of kicking, $a$ standing for a set of parameters which characterizes the 
oscillator.   Let us solve the above 
equation for $t_n-\delta \le t\le t_{n+1}-\delta $, $\delta $ being a 
positive infinitesimal quantity.  Then, taking the limit $\tau 
\rightarrow 0$, one finds that the above set of equations of motion reduces to
\begin{eqnarray}
\psi_{n+1}=e^{i\omega }f_a(\psi_n,\psi_n^*)
\end{eqnarray}
and $p(t_{n+1}-\delta )=i\omega \psi_{n+1}$, where we noticed that 
$\psi_n\equiv \psi (t_n-\delta )=\psi (t_n)$  because of the continuity 
of $\psi (t)$ at $t_n$.  
Here we defined
\begin{eqnarray}
f_a(\psi, \psi^*)=\psi +F_a(\psi ,\psi^*).
\end{eqnarray}
Equation ($2\cdot 5$) with eq.~($2\cdot 6$) is the mapping system for 
the kicked-oscillator 
system ($2\cdot 3$) and ($2\cdot 4$).  One should note that the mapping 
model ($2\cdot 5$) has two types 
of characteristic parameters.  One is the eigenfrequency $\omega $ of the 
oscillation and therefore is relevant to the phase degree of freedom, 
and the other is the parameter set $a$, which is regarded as the 
control parameter mainly relevant to the amplitude dynamics of oscillation.
\par
In the above, we proposed a mapping model relevant to a chaotic dynamics 
showing a well-defined oscillation.  Next, we will construct a 
mapping system consisted of coupled oscillators.  As an example, consider the 
two oscillators system coupled to each other,
\begin{eqnarray}
\dot{\psi }^{(j)}&=&p^{(j)},\\
\tau \dot{p}^{(j)}&=&i\omega_{j}\psi^{(j)}-p^{(j)}+F_{a_{j}}(\psi^{(j)},
\psi^{(j)*})\cdot \sum_{n=-\infty }^{+\infty }\delta (t-t_n) 
+\frac{K}{2}(\psi^{(k)}-\psi^{(j)})
\end{eqnarray}
for $(j,k)=(1,2),\ (2,1)$.  The last term of eq.~($2\cdot 8$) represents the 
coupling and $K$ is the coupling constant.  In the present paper, unless it is 
stated, $K$ is kept to be non-negative.  For the coupled $N$-elements system, the 
equations of motion for the $j$-th oscillator are written as
\begin{eqnarray}
\dot{\psi }^{(j)}&=&p^{(j)},\\
\tau \dot{p}^{(j)}&=&i\omega_j\psi^{(j)}-p^{(j)}+F_{a_j}(\psi^{(j)},
\psi^{(j)*})\cdot \sum_{n=-\infty }^{+\infty }\delta (t-t_n)+{\cal D}\psi^{(j)}.
\end{eqnarray}
Here ${\cal D}\psi^{(j)}$ represents the linear coupling term.  
Particularly, for the $N=2$ model given in eqs.~($2\cdot 7$) and ($2\cdot 8$), 
we get 
$
{\cal D}\psi^{(1,2)}=(K/2)(\psi^{(2,1)}-\psi^{(1,2)}).
$

Solving the above equations of motion for $t_n-\delta \le t\le t_{n+1}-\delta $ 
and taking the limit $\tau \rightarrow 0$, one obtains the coupled map system
\begin{eqnarray}
\psi_{n+1}^{(j)}&=&e^{i\omega_j+{\cal D}}f_{a_j}(\psi_n^{(j)},\psi_n^{(j)*}) 
\nonumber \\
&=&\sum_kJ_{jk}e^{i\omega_k}f_{a_k}(\psi_n^{(k)},\psi_n^{(k)*}),
\end{eqnarray}
for $\psi_n^{(j)}\equiv \psi^{(j)}(t_n-\delta )=\psi^{(j)}(t_n)$ and 
$p^{(j)}(t_{n+1}-\delta )=(i\omega_j+{\cal D}) \psi_{n+1}^{(j)}$, 
where $f_{a}(\psi ,\psi^*)$ is the same as in eq.~($2\cdot 6$).  
The coupling constant $J_{jk}$ has been defined as follows.  
Let us introduce the matrix $\hat{\Lambda }$ with the $jk$ element 
$\Lambda_{jk}$ defined by
\begin{eqnarray}
(i\omega_j+{\cal D})g_j=\sum_k\Lambda_{jk}g_k
\end{eqnarray}
for an arbitrary quantity $g_j$ defined for the oscillator $j$.  
The interaction kernel $J_{jk}$ in eq.~($2\cdot 11$) is defined via
\begin{eqnarray}
e^{i\omega_j +{\cal D}}g_j=\sum_k\left( e^{\hat{\Lambda }} \right)_{jk}g_k=
\sum_kJ_{jk}e^{i\omega_k}g_k.
\end{eqnarray}  
Equation ($2\cdot 11$) together with eqs.~($2\cdot 12$) and ($2\cdot 13$) 
is the fundamental result of the present paper.  

Particularly, if the coupling term has the form
\begin{eqnarray}
{\cal D}g_j=\sum_{k=1}^NK_{jk}( g_k-g_j),
\end{eqnarray}
where $K_{jk}$ is the coupling constant, 
then the matrix element is given by
\begin{eqnarray}
\Lambda_{jk}= i\omega_j\delta_{jk}+K_{jk}
-\left( \sum_{\ell =1}^NK_{j\ell }\right) \delta_{jk} .
\end{eqnarray}
\par
If all characteristic frequencies are same, i.e., 
$\omega_1=\omega_2=\cdots =\omega_N\equiv \omega $, and 
the coupling operator satisfies the condition
$
\sum_j{\cal D}g_j=0
$ 
and therefore $\sum_{j}J_{jk}=1$, then eq.~($2\cdot 11$) is rewritten as
\begin{eqnarray}
\psi_{n+1}^{(j)}=e^{i\omega }\left\{ f_{a_j}(\psi_n^{(j)},\psi_n^{(j)*})
+\sum_kJ_{jk}\left[ f_{a_k}(\psi_n^{(k)},\psi_n^{(k)*})
-f_{a_j}(\psi_n^{(j)},\psi_n^{(j)*})\right] \right\} .
\end{eqnarray}
Furthermore, for the global coupling 
$
{\cal D}g_j=(K/N)\sum_{k=1}^N( g_k-g_j),
$
one obtains $J_{jk}=K/N$ for any combinations of $j$ and $k$.  
In addition, when the system characteristics are all the same ($a_1=a_2=\cdots 
=a_N\equiv a$), one finds that the equations of motion under consideration 
has the complete synchronization $\psi_n^{(1)}=\psi_n^{(2)}=\cdots 
=\psi_n^{(N)}\equiv \psi_n^0$, which obeys
$
\psi_{n+1}^0=e^{i\omega }f_a(\psi_n^0,\psi_n^{0*})
$
irrespectively of the coupling strength.  This 
fact suggests the possibility of the existence of the transition between 
the complete synchronization and its broken state.  Details will be 
given in Sec.~3 for the two-oscillators system.
\section{Chaotic phase synchronization in a two oscillators system}
In this section, we study the synchronization and desynchronization for 
the coupled maps model of a two oscillators system proposed in the 
preceding section.  For the 
coupling term ${\cal D}g_{1,2}=(K/2)(g_{2,1}-g_{1,2})$, we get
\begin{eqnarray}
\hat{\Lambda }=
\left(
\begin{array}{cc}
i\omega_1-\frac{K}{2}  &  \frac{K}{2}  \\
\frac{K}{2} & i\omega_2-\frac{K}{2}
\end{array}
\right) .
\end{eqnarray}
This immediately gives the coupled map system
\begin{eqnarray}
\psi_{n+1}^{(1)}=J_K(\omega_1-\omega_2)
e^{i\omega_1} f_{a_1}(\psi_n^{(1)},\psi_n^{(1)*})+J_K'(\omega_1-\omega_2)
e^{i\omega_2}f_{a_2}(\psi_n^{(2)},\psi_n^{(2)*}) ,
 \\
\psi_{n+1}^{(2)}=
J_K'(\omega_2-\omega_1)
e^{i\omega_1} f_{a_1}(\psi_n^{(1)},\psi_n^{(1)*})+J_K(\omega_2-\omega_1)
e^{i\omega_2}f_{a_2}(\psi_n^{(2)},\psi_n^{(2)*}) ,
\end{eqnarray}
where the coupling constants are
\begin{eqnarray}
J_K(\Delta \omega )&=&e^{-\frac{i}{2}\Delta \omega }e^{-\frac{K}{2}}\left[ 
\cosh \frac{\sqrt{K^2-(\Delta \omega)^2}}{2}+i\Delta \omega \frac{\sinh 
\frac{\sqrt{K^2-(\Delta \omega)^2}}{2}}{\sqrt{K^2-(\Delta \omega)^2}} \right] ,
\\ 
J_K'(\Delta \omega )&=&e^{\frac{i}{2}\Delta \omega }e^{-\frac{K}{2}}
\frac{K\sinh \frac{\sqrt{K^2-(\Delta \omega)^2}}{2}}{\sqrt{K^2-(\Delta \omega)^2}} .
\end{eqnarray}
It should be noted that the coupling constants satisfy 
$J_K(\Delta \omega)^*=J_K(-\Delta \omega)$ 
and $J_K'(\Delta \omega)^*=J_K'(-\Delta \omega)$.
\subsection{General characteristics of the two oscillators system}
Since $J_0(\Delta \omega )=1$ and $J_0'(\Delta \omega )=0$, 
oscillators are independent of each other for $K=0$.  On the other 
hand, for $K\rightarrow \infty $, noting $J_{\infty } (\Delta \omega )=
\frac{1}{2}e^{-\frac{i}{2}\Delta \omega }=J_{\infty}'(\Delta \omega )^*$, 
we obtain
\begin{eqnarray}
\psi_{n}^{(1)}&=&\psi_{n}^{(2)}\equiv \psi_{n}^{0},\\
\psi_{n+1}^0&=&\frac{1}{2}e^{\frac{i}{2}(\omega_1+\omega_2 )}\left[ f_{a_1}
(\psi_n^0,\psi_n^{0*})+f_{a_2}(\psi_n^0,\psi_n^{0*}) \right]  .
\end{eqnarray}
This result implies that even if the characteristics of two oscillators are 
different, the system shows a complete synchronization in the strong 
coupling limit $K\rightarrow \infty$.  It should be noted that 
for the coupled system consisted of identical oscillators the complete 
synchronization is achieved irrespectively of the strength as a particular 
motion, while the present synchronization for non-identical oscillators 
is observed only when the $K\rightarrow \infty $ limit.\footnote{
The above result holds for $K>0$.  In the opposite limit ($K\rightarrow 
-\infty $), we observe the ``anti-phase'' oscillation $\psi_n^{(1)}=
-\psi_n^{(2)}$.  Namely for $K\rightarrow -\infty $, noting $J_K(\Delta 
\omega )=\frac{1}{2}e^{-\frac{i}{2}\Delta \omega +|K|},\ J_K'(\Delta 
\omega )=-J_K(\Delta \omega )^*$, one gets the equation of motion
$$
\psi_{n+1}^{(1)}=-\psi_{n+1}^{(2)}=\frac{1}{2}e^{\frac{i}{2}(\omega_1+
\omega_2)+|K|}\left[ f_{a_1}(\psi_n^{(1)},\psi_n^{(1)*})- f_{a_2}
(-\psi_n^{(1)},-\psi_n^{(1)*})\right] .
$$
}  
On the other hand, if the system parameters $\omega_j$ and $a_j$ are different 
from each other, the system does not show a complete synchronization.  
Nevertheless, we expect that there exists a certain kind of synchronization 
state if the coupling constant $K$ is suitably chosen, 
which is nothing but the phase synchronization~\cite{RPK96}.  
This will be studied in the following subsection. 
\par
Particularly, if $\omega_1=\omega_2\equiv \omega $ and $a_1=a_2\equiv a$, 
i.~e.~, the 
two oscillators are identical, then the coupled map system is written as
\begin{eqnarray}
\psi_{n+1}^{(1)}&=&e^{i\omega }\left\{ f_{a}(\psi_n^{(1)},\psi_n^{(1)*})
+J_{k}'\left[ f_{a}(\psi_n^{(2)},\psi_n^{(2)*})
-f_{a}(\psi_n^{(1)},\psi_n^{(1)*})\right] \right\},\\
\psi_{n+1}^{(2)}&=&e^{i\omega }\left\{ f_{a}(\psi_n^{(2)},\psi_n^{(2)*})
+J_{k}'\left[ f_{a}(\psi_n^{(1)},\psi_n^{(1)*})
-f_{a}(\psi_n^{(2)},\psi_n^{(2)*})\right] \right\},
\end{eqnarray}
where $J_K'=(1-e^{-K})/2$.  In this case, the particular, {\it complete 
synchronization} state exists and the dynamical variable $\psi_n^0(
=\psi_n^{(1)}=\psi_n^{(2)})$ obeys $\psi_{n+1}^0=e^{i\omega }f_a(\psi_n^0,
\psi_n^{0*})$.  Around the complete synchronization state, the small 
deviations $\delta \psi_n^{(1)}=\psi_n^{(1)}-\psi_n^0$ and $\delta 
\psi_n^{(2)}=\psi_n^{(2)}-\psi_n^0$ from the complete synchronization obey
\begin{eqnarray}
\delta \psi_{n+1}^+&=&e^{i\omega }(G_{1,n}\delta \psi_n^+ +G_{2,n}\delta 
\psi_n^{+*}),\\
\delta \psi_{n+1}^-&=&e^{i\omega }e^{-K}(G_{1,n}\delta \psi_n^- +G_{2,n}
\delta \psi_n^{-*}),\end{eqnarray}
where $\delta \psi_n^{\pm }=(\delta \psi_n^{(1)}\pm \delta \psi_n^{(2)} )/2$, 
and $G_{1,n}=\partial f_a(\psi_n^0,\psi_n^{0*})/\partial \psi_n^0$ and 
$G_{2,n}=\partial f_a(\psi_n^0,\psi_n^{0*})/\partial \psi_n^{0*}$.  It 
should be noted that the two equations of motion for $\delta \psi_n^{\pm }$ 
are separated from each other, and the 
first equation is identical to the perturbation equation $\delta \psi_{n+1}^0
=e^{i\omega }(G_{1,n}\delta \psi_n^0 +G_{2,n}\delta \psi_n^{0*})$ for the 
small change of the initial condition of the element dynamics.  Let 
$\lambda $ be the largest Lyapunov exponent of the element dynamics.  One 
finds that
\begin{eqnarray}
|\delta \psi_n^+|\sim e^{\lambda n}|\delta \psi_0^+|
\end{eqnarray}
for large $n$.  The exponent $\lambda $ shows the characteristic of the 
uncoupled system and is free from the stability of the complete 
synchronization.  On the other hand, the second equation gives
\begin{eqnarray}
|\delta \psi_n^-|=\frac{1}{2}|\psi_n^{(2)}-\psi_n^{(1)}|
\sim e^{\lambda_{\perp } n}|\delta \psi_0^-|,
\end{eqnarray}
where
\begin{eqnarray}
\lambda_{\perp }=\lambda -K.
\end{eqnarray}
The parameter $\lambda_{\perp }$ called the {\it transverse Lyapunov exponent} 
or the {\it stability parameter} determines the stability of the 
complete synchronization~\cite{FY83}.  Namely, if $\lambda_{\perp } <0$, 
the complete 
synchronization is linearly stable.  On the other hand, if 
$\lambda_{\perp }>0$, the complete synchronization is unstable.  
If the element dynamics is periodic, then $\lambda <0$ and therefore 
$\lambda_{\perp}<0$, which implies that the complete synchronization 
is linearly stable.  On the other hand, for a chaotic element 
dynamics ($\lambda >0$), 
the system shows the transition at $K=\lambda (\equiv K_c)$ as the 
coupling strength $K$ is decreased from above $K_c$, and one observes 
the on-off (modulated) intermittency for $K$ slightly 
below $K_c$~\cite{FY85,FY86}.
\subsection{Chaotic phase synchronization and its breakdown}
In the present paper, as a model of element dynamics, we use the mapping function
\begin{eqnarray}
f_{a}(\psi ,\psi^*)=[a-(1+ib)|\psi |^2]\psi ,
\end{eqnarray}
where the parameters $a$ and $b$ are chosen in such a way that the 
dynamics $\psi_{n+1}^{(j)}=e^{i\omega_j}f_{a_j}(\psi_n^{(j)},\psi_n^{(j)*})$ 
shows a chaotic behavior.  One should note that the statistical 
characteristic of the isolated element dynamics is free from the choice 
of $\omega_j$ 
because the local dynamics does not depend on the eigenfrequency after the 
replacement $\psi_n^{(j)}\rightarrow e^{i\omega_jn}\psi_n^{(j)}$.  
A trajectory of the element dynamics ($3\cdot 15$) for $\omega =0.08, a=2.55$ and 
$b=0$ is shown in Fig.~1.  Figure 2 depicts how the trajectories depend on 
$b$.  One observes that depending on $b$, the isolated dynamics shows 
various characteristics.  In later numerical experiments of coupled map system, 
we use the same values 
for $a$ and $b$ as $a_1=a_2=2.55$ and $b_1=b_2=0$ for which the element dynamics 
are chaotic, where the largest Lyapunov exponent $\lambda $ is about 0.144, 
and the frequency parameters $\omega_1$ and $\omega_2$ are chosen to be set as 
$\Delta \omega =\omega_1-\omega_2=0.08$ except the figure 4.  One should note that 
in the present model only the difference of eigenfrequencies is relevant.  
It was numerically proved that as far as $b$ is small enough, 
the introduction of non-vanishing $b_j$ does not change the qualitative 
results found for $b_j=0$.
\begin{wrapfigure}{l}{6.6cm}
  \centerline{\includegraphics[angle=-90, width=60mm]{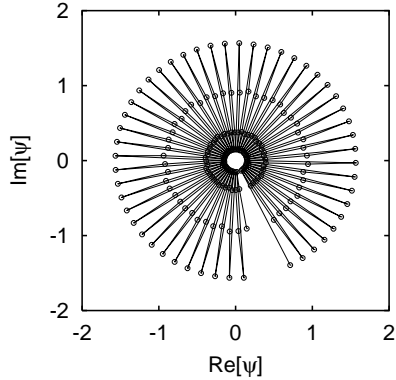}}
\caption{
Trajectory for the element dynamics ($2\cdot 5$) 
with the mapping function ($3\cdot 15$), where 
$\omega =0.03, a=2.55$ and $b=0$.  The dynamics is chaotic 
with the largest Lyapunov exponent $\lambda =0.144$.}
\label{fig:1}
\end{wrapfigure}
  
Let $\theta_n^{(j)}$ be the phase of $\psi_n^{(j)}$, i.e., 
$\psi_n^{(j)}=|\psi_n^{(j)}|e^{i\theta_n^{(j)}}$.  It should be noted 
that as far as the 
phase variation is small enough, the phase dynamics of $\theta_n^{(j)}$ 
is uniquely determined even for a coupled system.  For the 
control parameters $a_j$ and $b_j$ used in the present paper, 
the phase variations are small enough.  
The phase difference $\Delta \theta_n=\theta_n^{(1)}-
\theta_n^{(2)}$ obeys the equation of motion
\begin{eqnarray}
\Delta \theta_{n+1}=\Delta \theta_n+\alpha_n (\Delta \theta_n),
\end{eqnarray}
where $\alpha_n (\Delta \theta )$, which is uniquely determined by 
eqs.~($3\cdot 2$) and ($3\cdot 3$), is a $2\pi $-periodic function of 
$\Delta \theta $, i.e., 
$(\alpha_n (\Delta \theta +2\pi )=
 \alpha_n (\Delta \theta ))$.  The integration of the above equation yields
\begin{eqnarray}
\Delta \theta_n =\Delta \theta_0+\sum_{m=0}^{n-1}\alpha_m (\Delta \theta_m ).
\end{eqnarray}
By taking the statistical average of the above equation, which is equivalent 
to the time average with the ergodicity assumption, the angular velocity of 
the phase difference (rotation number) is given by
\begin{eqnarray}
\Delta \Omega  =\lim_{n\rightarrow \infty }
\frac{1}{n}\sum_{m=0}^{n-1}\alpha_m (\Delta \theta_m ) =\langle \alpha \rangle , 
\end{eqnarray}
where $\langle \cdots \rangle $ is the statistical average over an ensemble 
representing the steady state of the present chaos.  
Therefore, the average value of the phase difference obeys
\begin{eqnarray}
\langle \Delta \theta_n \rangle =\Delta \theta_0+\Delta \Omega \cdot n.
\end{eqnarray}

\begin{figure}[htb]
  \centerline{
  \includegraphics[angle=-90, width=40mm]{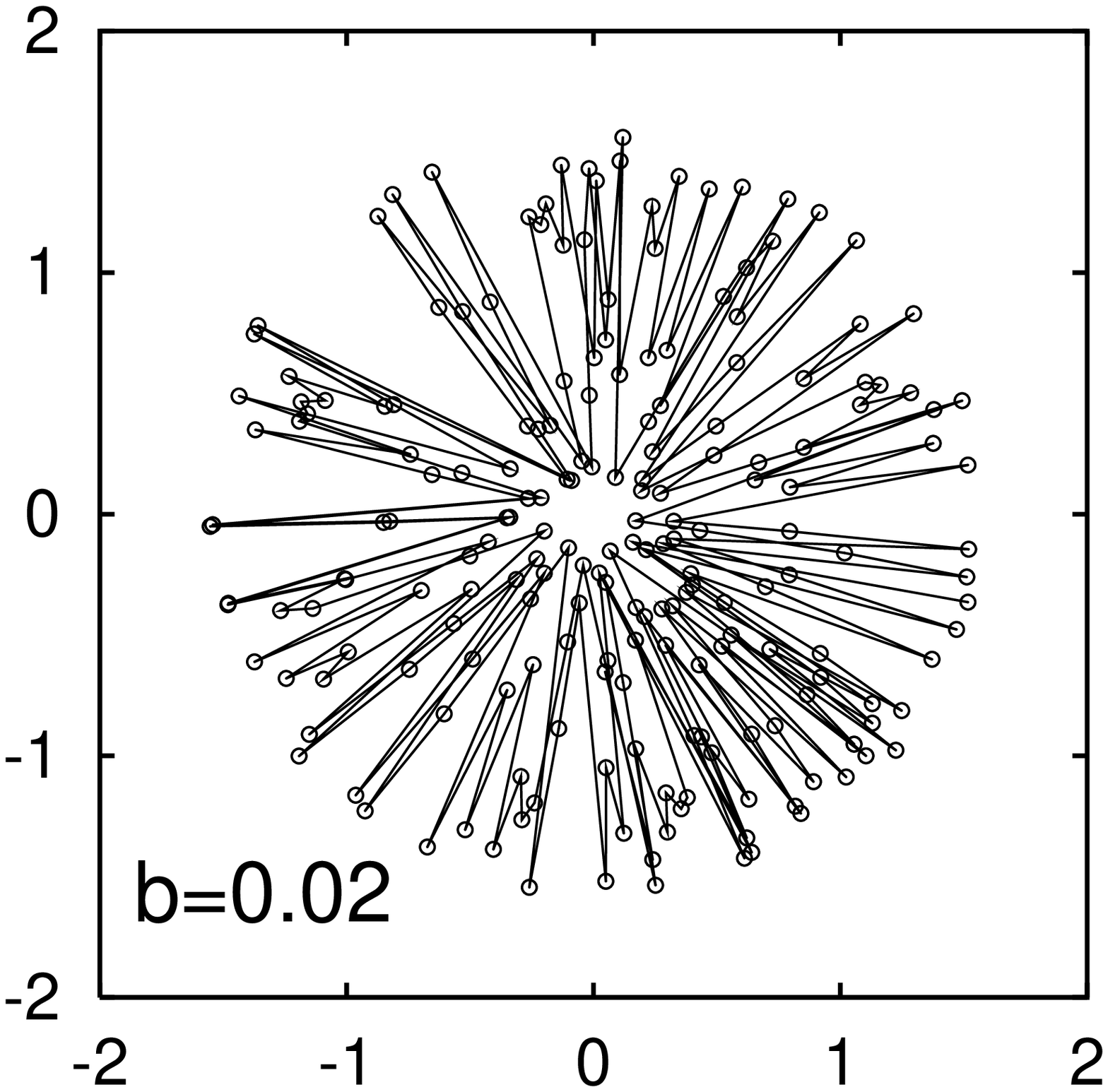}\ 
  \includegraphics[angle=-90, width=40mm]{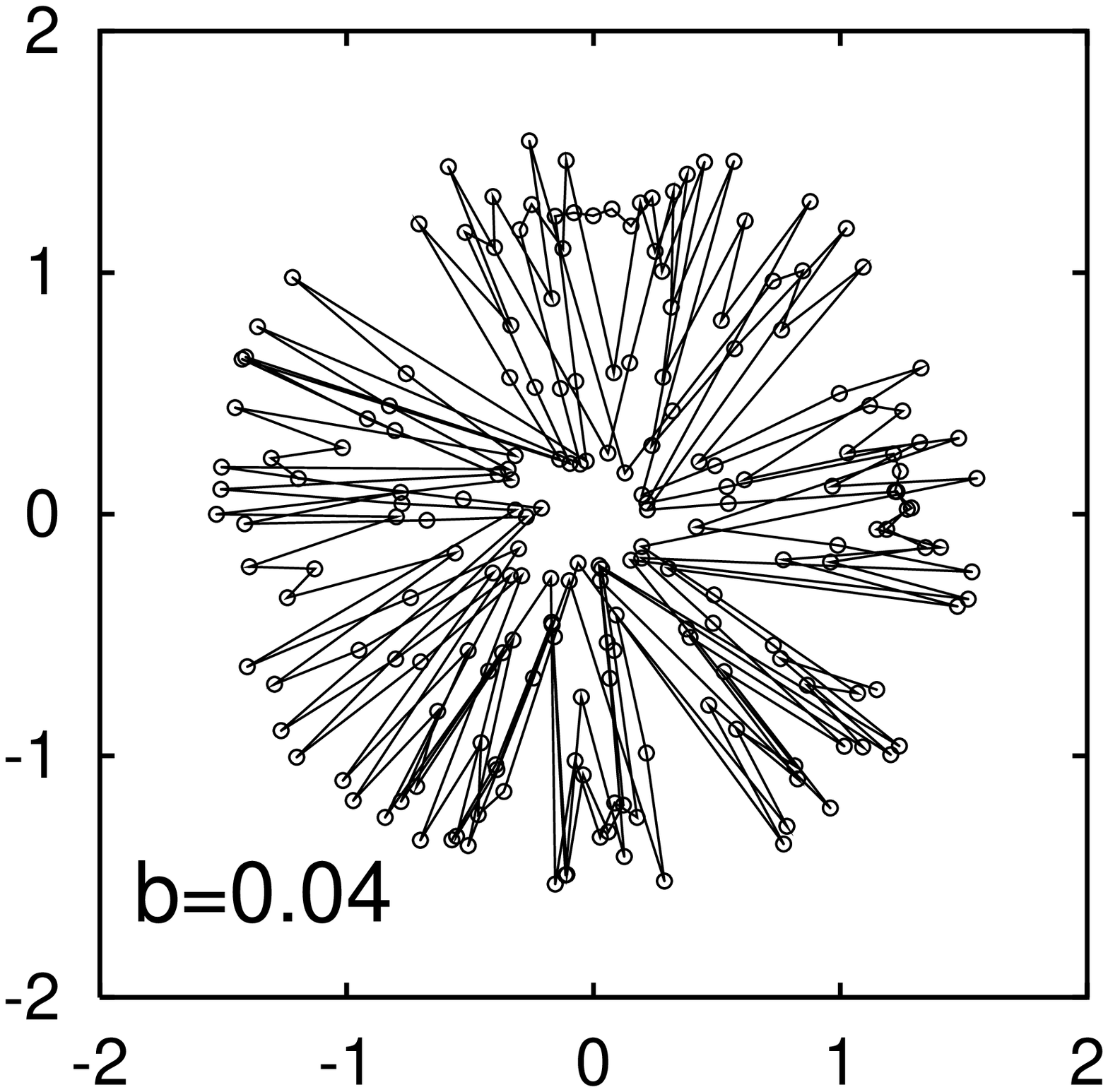}\ 
  \includegraphics[angle=-90, width=40mm]{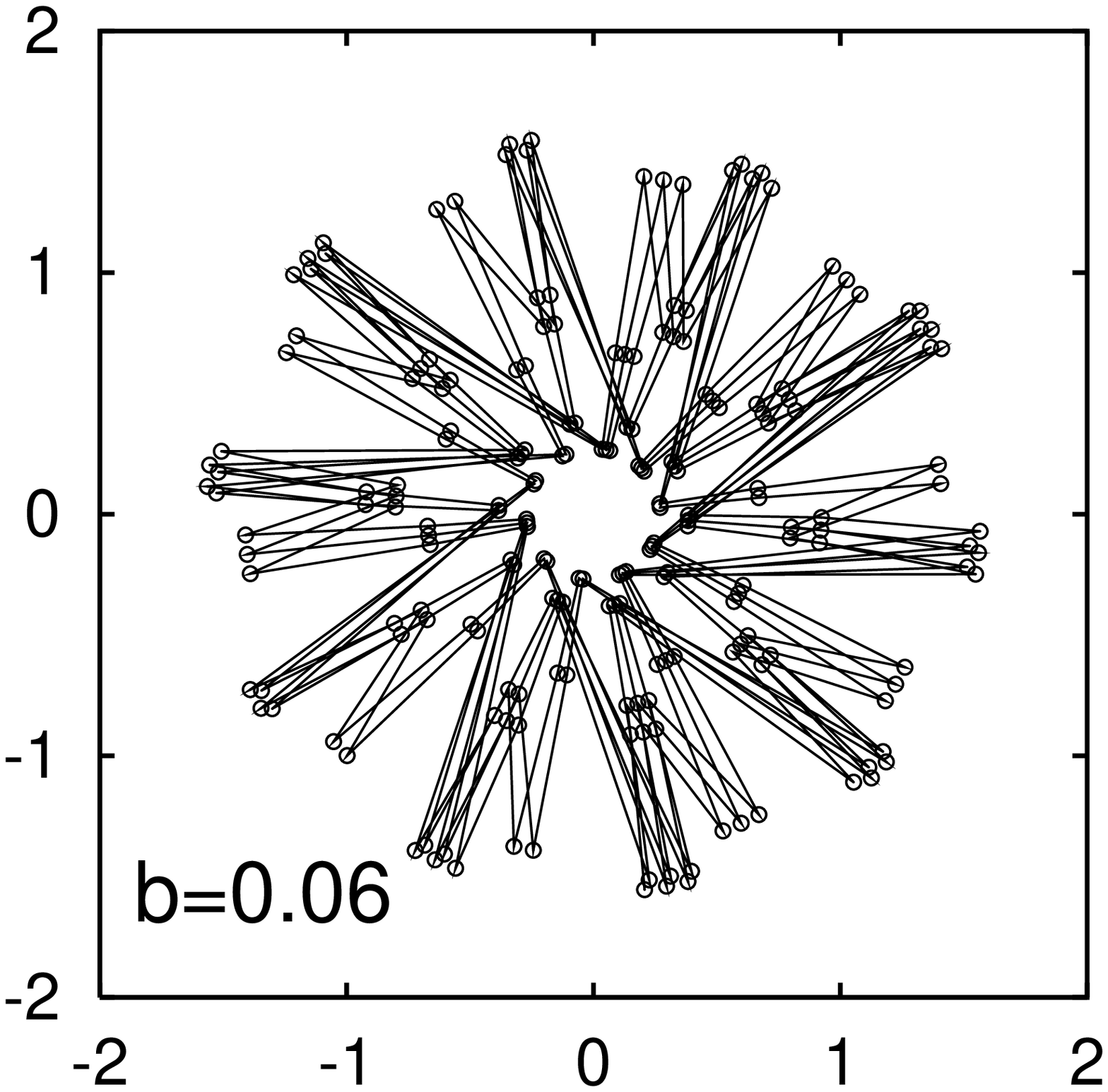}}
\centerline{
  \includegraphics[angle=-90, width=40mm]{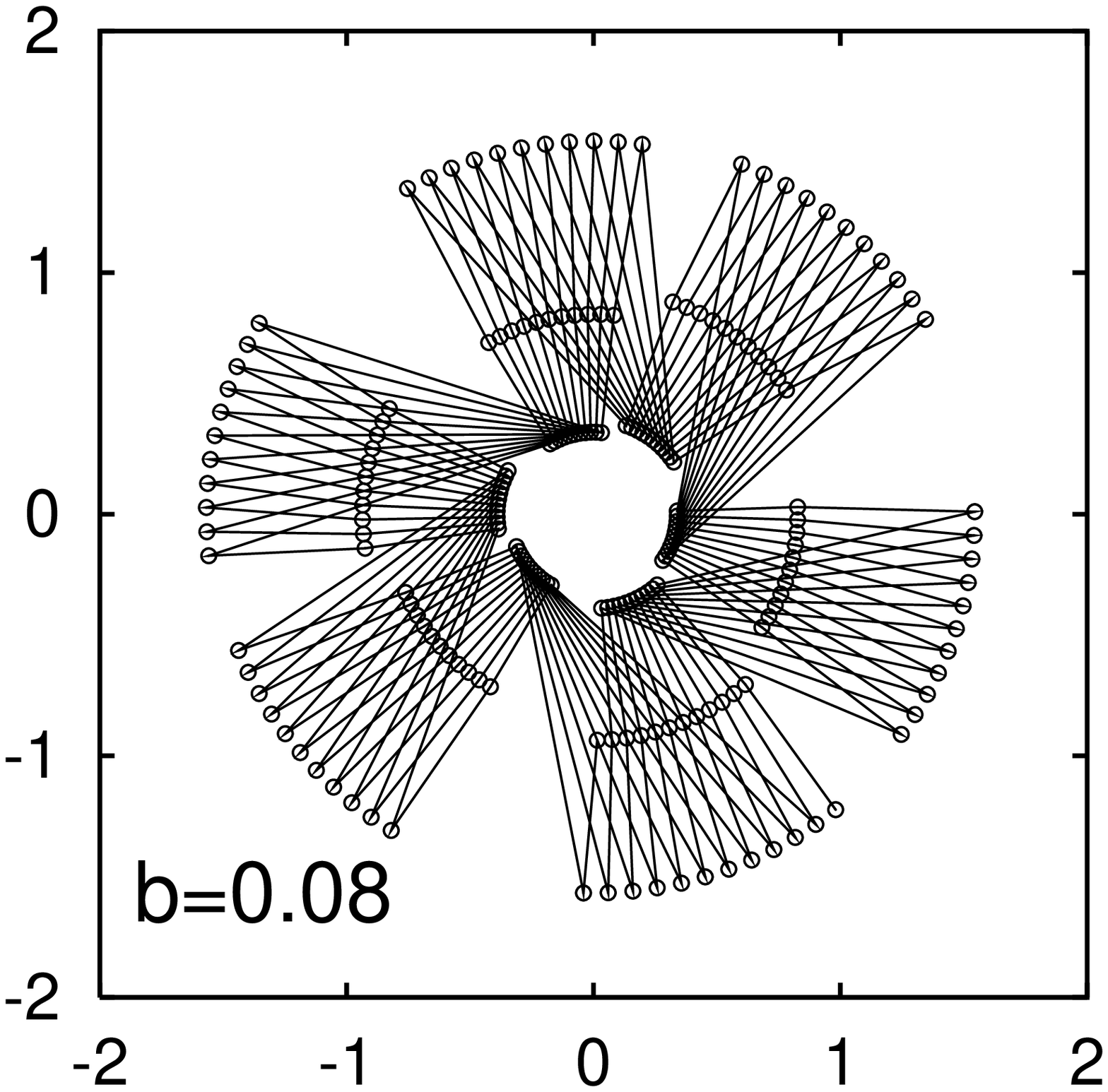}\ 
  \includegraphics[angle=-90, width=40mm]{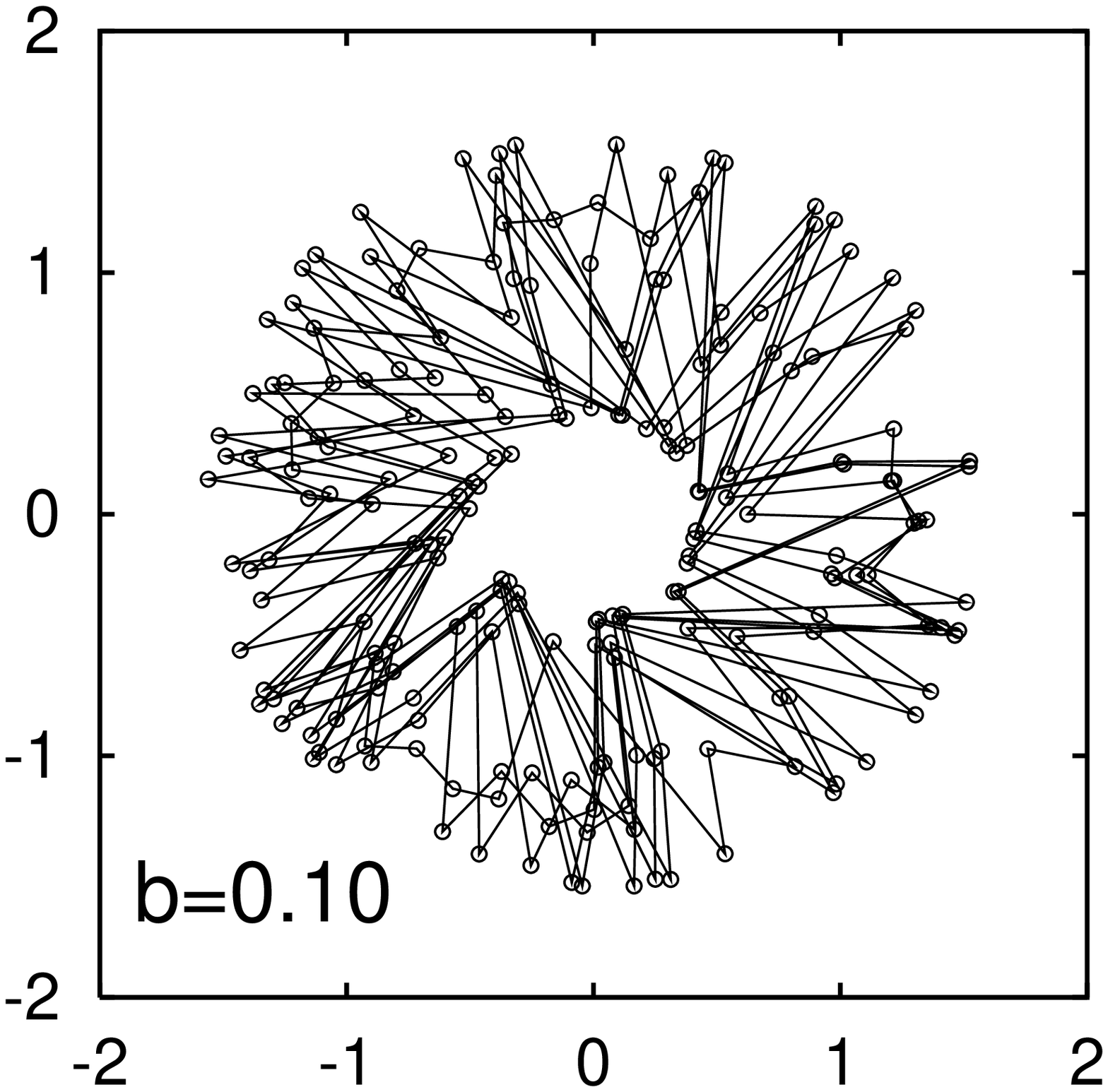}\ 
  \includegraphics[angle=-90, width=40mm]{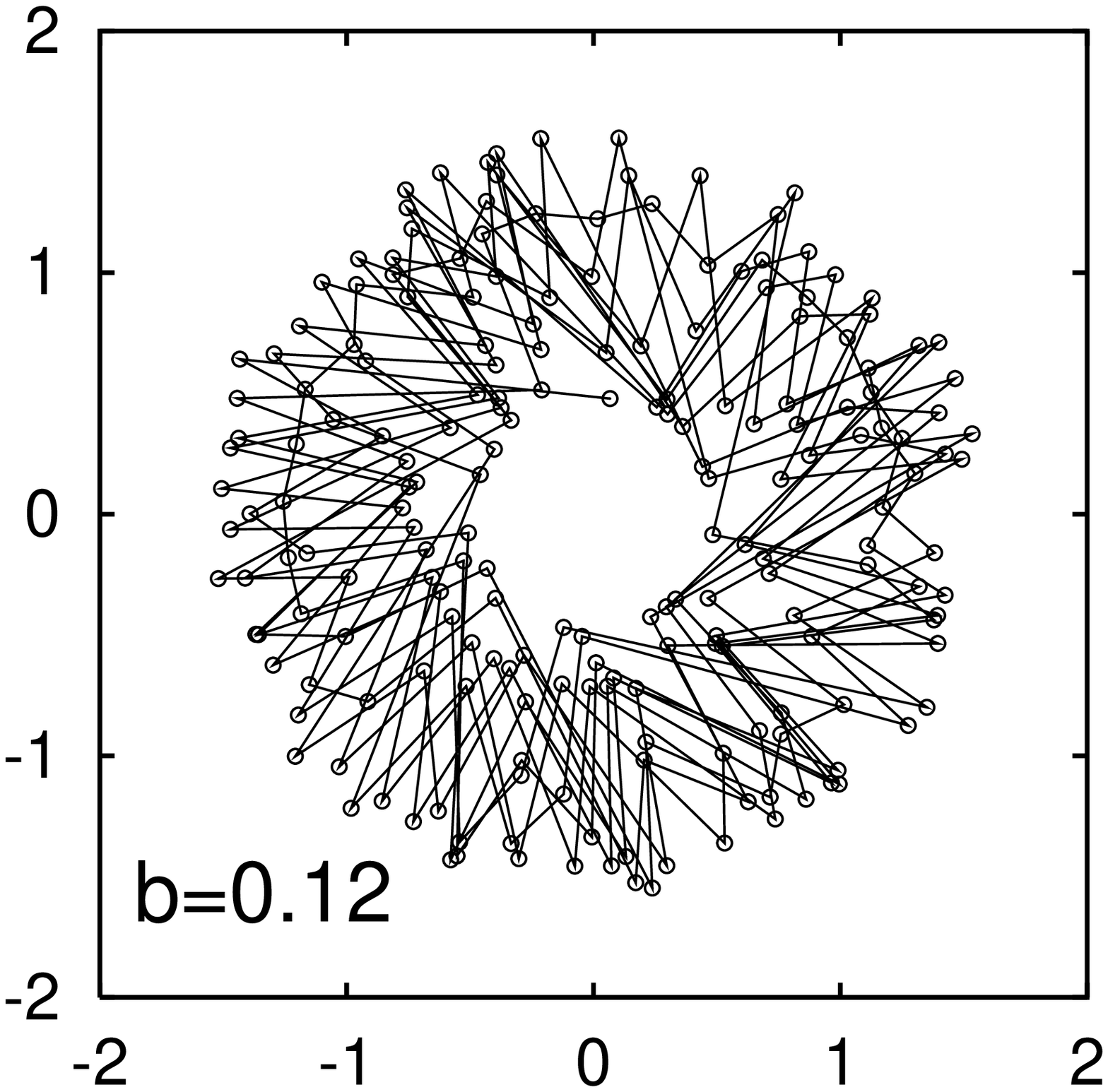}}
\centerline{
  \includegraphics[angle=-90, width=40mm]{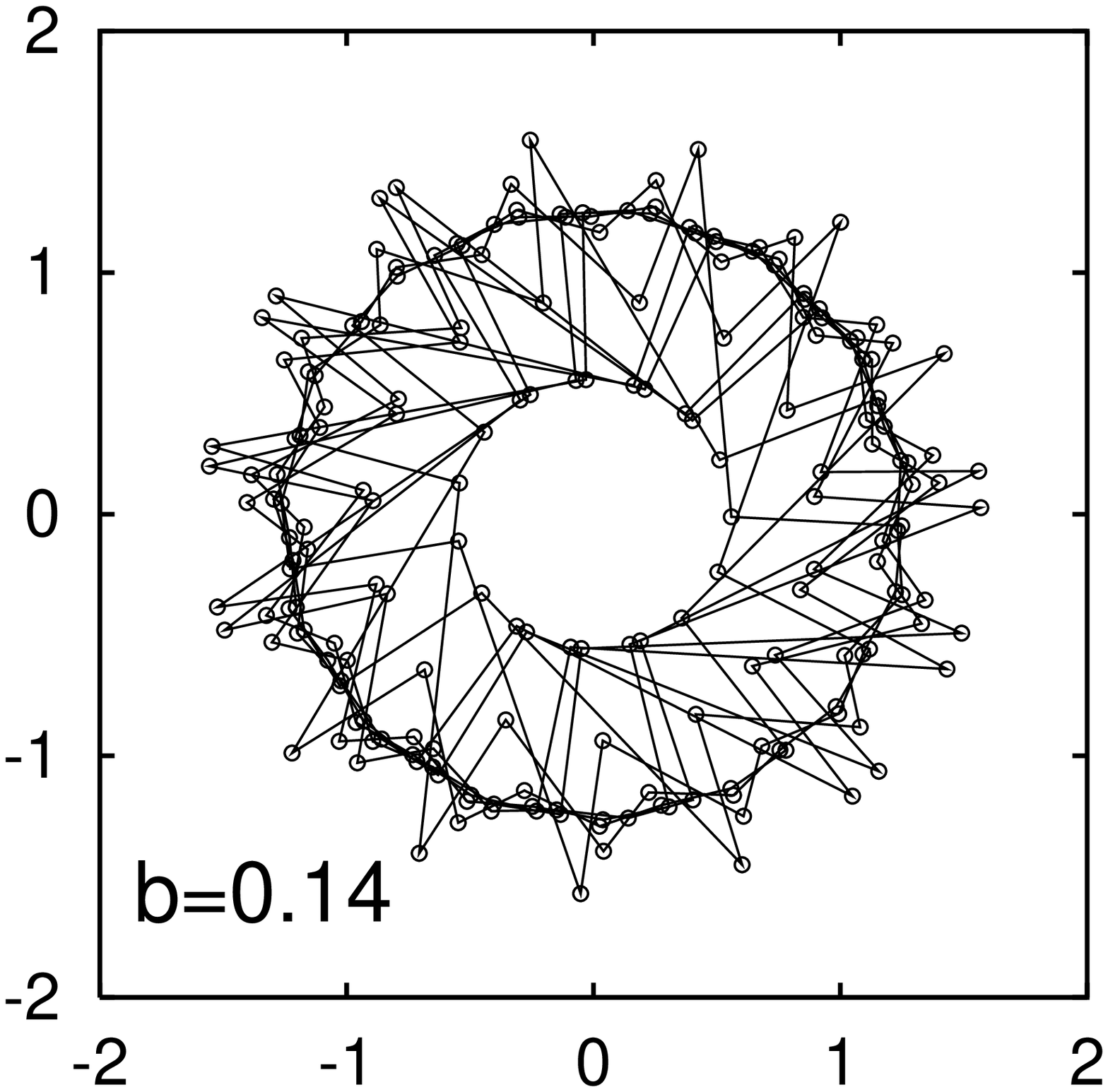}\ 
  \includegraphics[angle=-90, width=40mm]{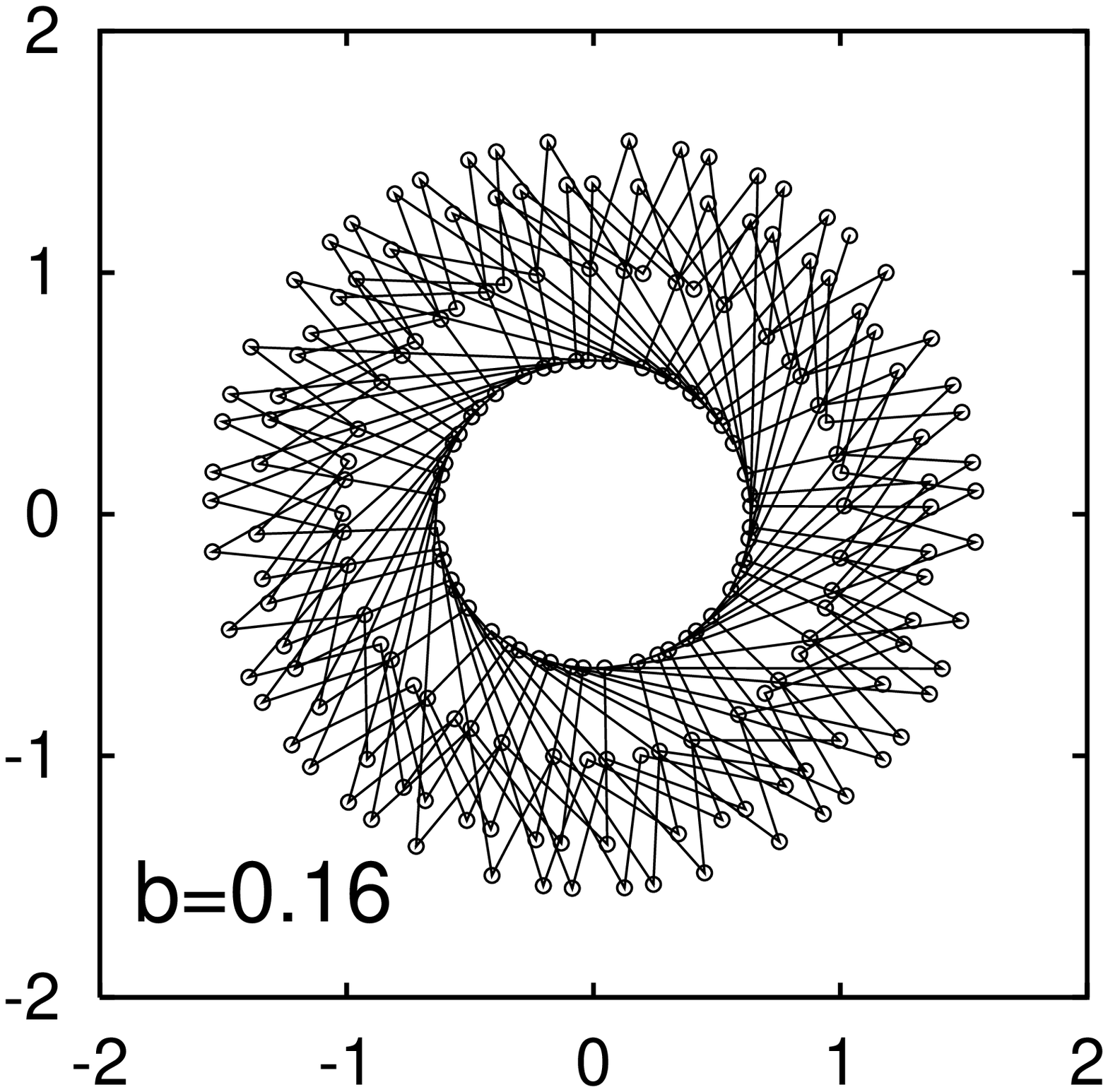}\ 
  \includegraphics[angle=-90, width=40mm]{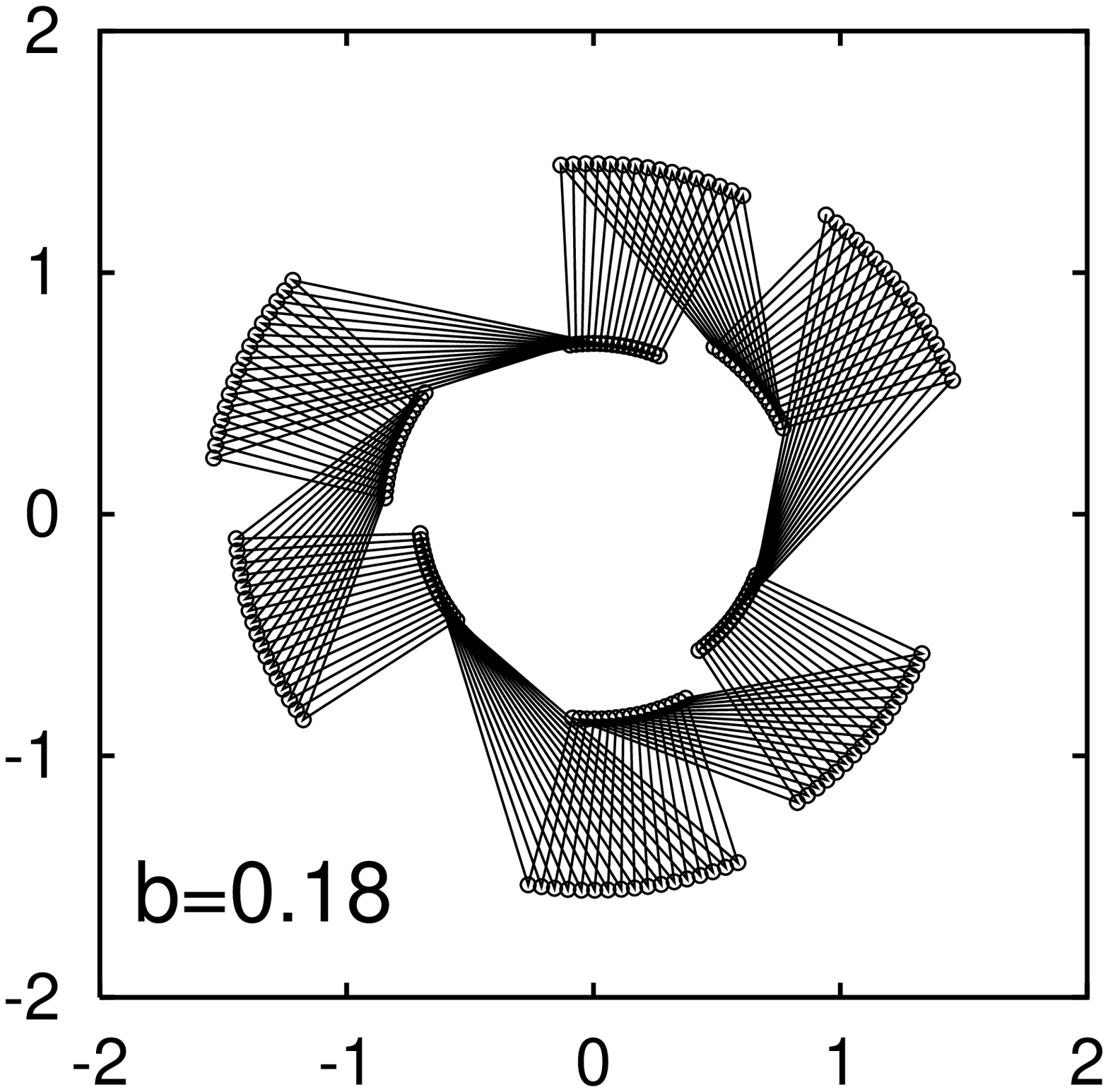}}
\caption{
Trajectories for the element dynamics ($2\cdot 5$) with the mapping 
function ($3\cdot 15$) for $a=2.55$ and various values of $b$.  
The abscissa and the ordinate are respectively 
${\rm Re}(e^{-i\omega }\psi_n)$ and ${\rm Im}(e^{-i\omega }\psi_n)$.  
Trajectories do not depend on the value of $\omega $ for the 
mapping ($3\cdot 15$).
}
\label{fig:2}
\end{figure}

Figure 3 shows temporal evolutions of the phase difference 
for different values of the coupling 
strength $K$.  One observes that depending on $K$, the coupled mapping system 
exhibits chaotic phase synchronization ($\Delta \Omega =0$).  
Figure 4 displays 
how the rotation number $\Delta \Omega $ depends on $K$ 
and $\Delta \omega (=\omega_1-\omega_2)$.  One finds that 
under the change of either $K$ or $\Delta \omega $ a clear transition 
between phase synchronization and phase desynchronization state occurs.
\begin{figure}[htb]
\parbox{\halftext}{
  \centerline{\includegraphics[angle=-90, width=60mm]{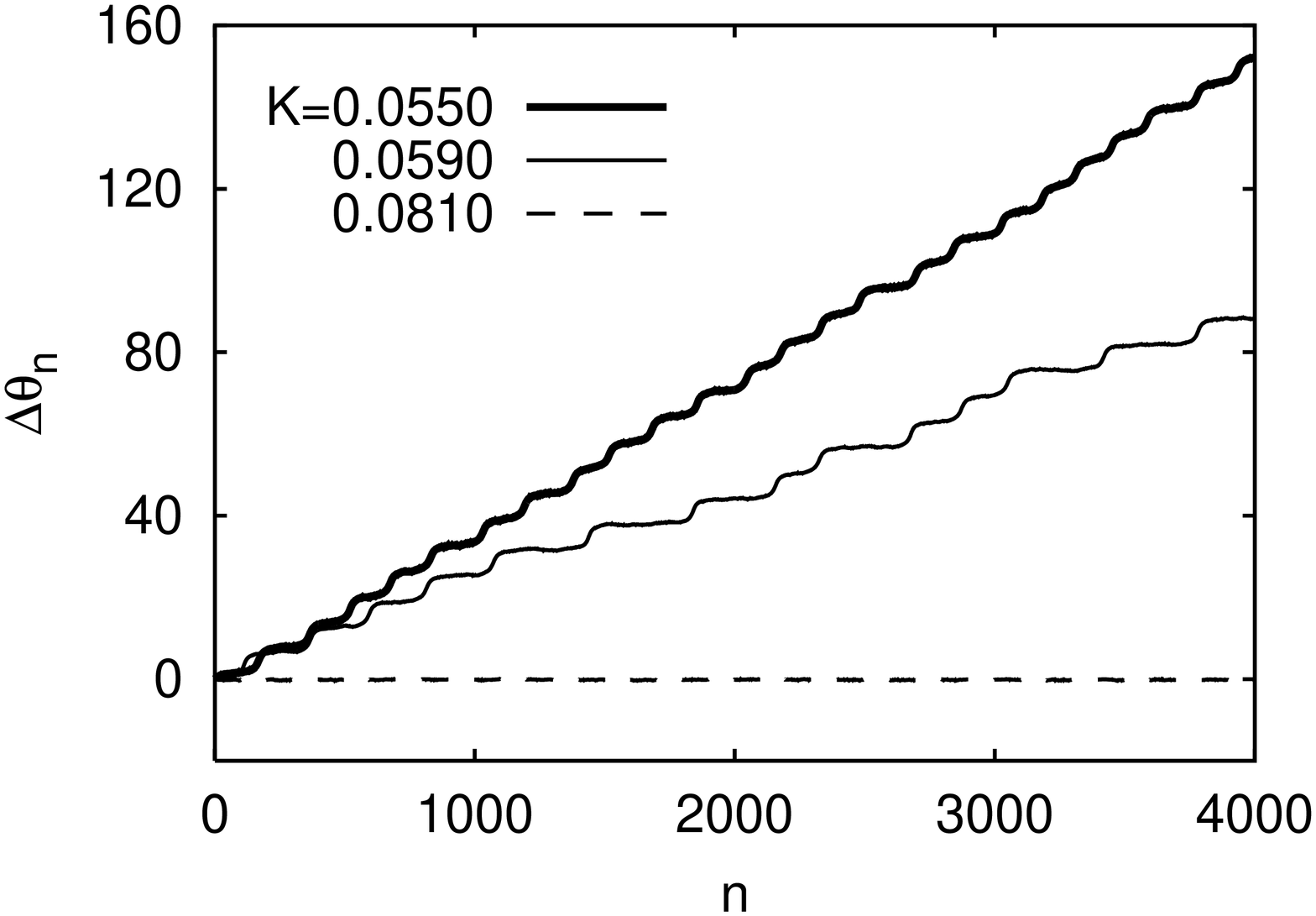}}
\caption{
Temporal evolution of phase differences for the 
coupled two maps system.  For $K=0.081$, the phase synchronization 
is achieved.
}
\label{fig:3}}
\hfill
\parbox{\halftext}{
  \centerline{\includegraphics[width=60mm]{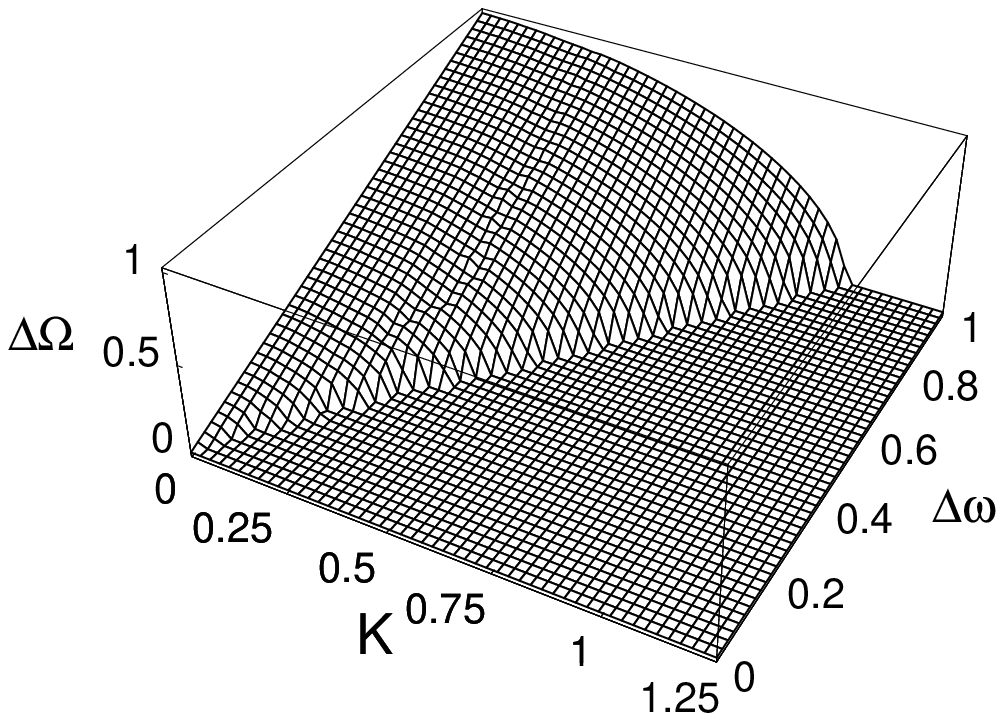}}
\caption{
The rotation number $\Delta \Omega $ depends on 
both the eigenfrequency mismatch $\Delta \omega (=\omega_1-\omega_2)$ 
and the coupling constant $K$.  For $\Delta \Omega =0$, the phase 
synchronization is achieved.   
}
\label{fig:4}}
\end{figure}
\par
In order to study the critical behavior of $\Delta \Omega $ as the 
coupling strength $K$ is decreased by keeping $\Delta \omega $ constant to be 
0.08, the quantity $(\Delta \Omega )^2$ is plotted as a function of $K$ in 
Fig.~5(a).  The figure suggests that the critical behavior
\begin{eqnarray}
\Delta \Omega \sim (K_*-K)^{\frac{1}{2}}
\end{eqnarray}
holds in the region $K<K_*$, where $K_*(\approx 0.0612)$ is a characteristic 
value of $K$ being below $K_c(\approx 0.0804)$, the desynchronization point, i.e., 
$\Delta \Omega =0$ for $K>K_c$.  The critical behavior $(3\cdot 20)$ is 
identical to 
that of the saddle-node bifurcation, and is well-known in association 
with the existence of the type I intermittency.  
Figure 5(b) shows the numerical result for the $K$ region 
below the onset point $K_c$ of the chaotic phase synchronization.  
Slightly below $K_c$, 
the rotation number turns out to take the asymptotic form
\begin{eqnarray}
\Delta \Omega \sim \exp \left[ -c\cdot (K_c-K)^{-\frac{1}{2}} \right] ,
\end{eqnarray}
instead of the saddle-node type, where $c$ is a positive constant.  
This behavior is a result of chaotic, one may say, ``stochastic" 
emergence of the saddle-node type channel, and has been reported near 
the desynchronization point in a phenomenological mapping model~\cite{POR97} 
and experimentally in the modulated CO$_2$ laser system~\cite{BAM02}.  
So, the present model belongs to the same universality class of chaotic 
phase synchronization showing the above two different critical behaviors.
\begin{figure}[htb]
  \centerline{
    \includegraphics[angle=-90, width=60mm]{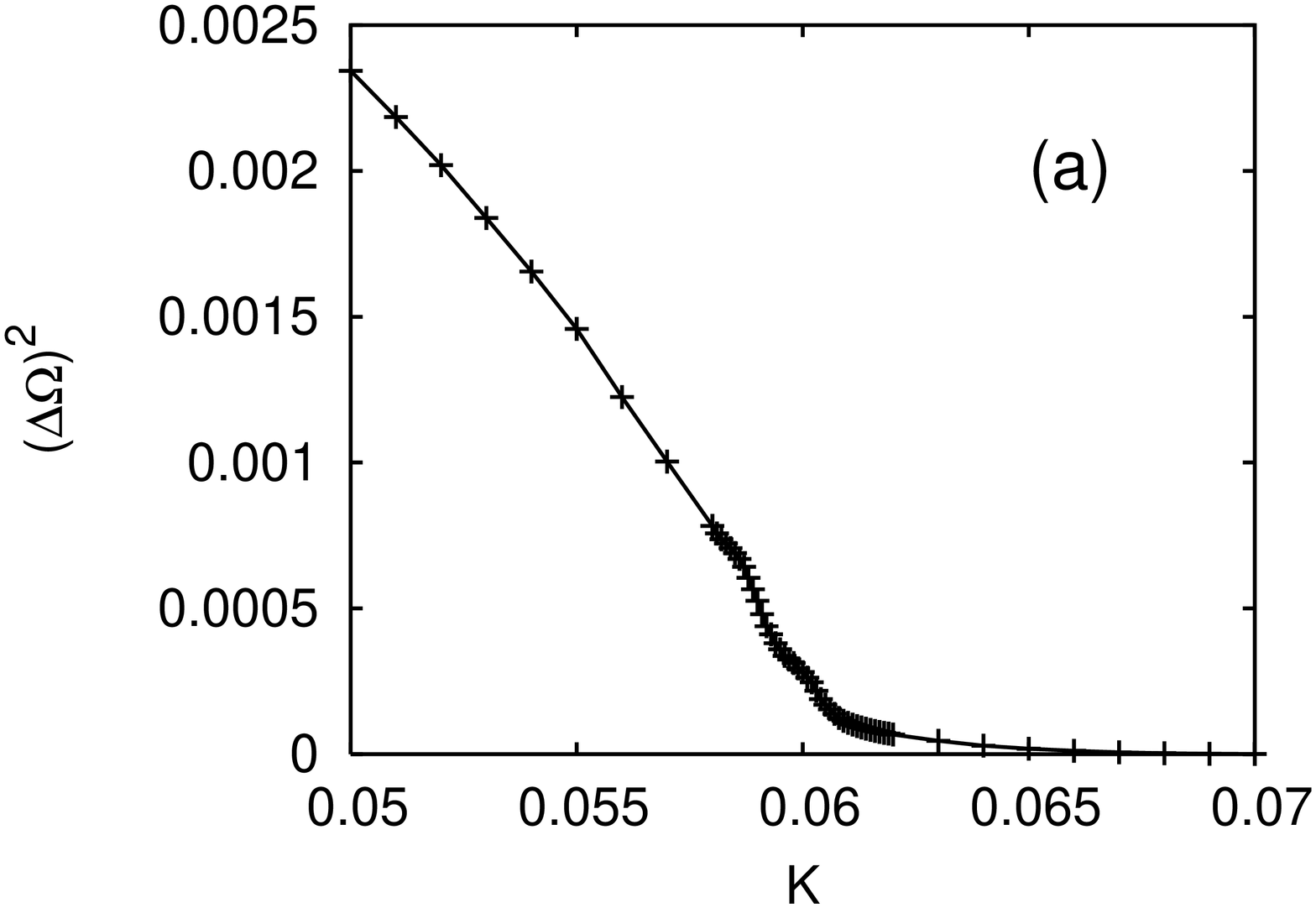}\quad
    \includegraphics[angle=-90, width=60mm]{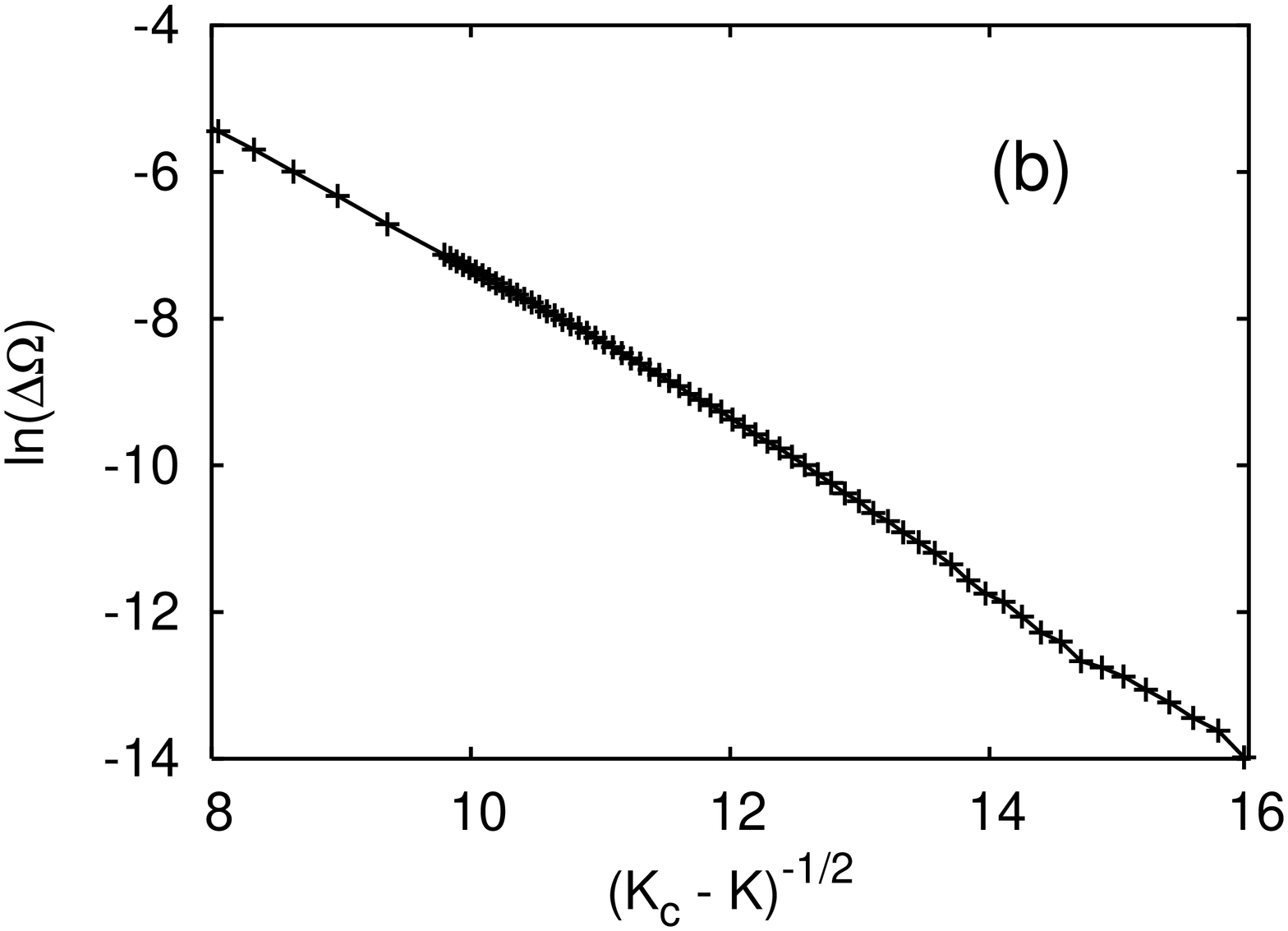}
  }
\caption{
Critical behaviors of the rotation number $\Delta \Omega $ as 
functions of the coupling constant $K$.
}
\label{fig:5}
\end{figure}
\par
The fluctuation of the phase difference $\Delta \theta_n$ from the 
average value ($3\cdot 19$) is evaluated with the variance
\begin{eqnarray}
\sigma_n^2\equiv \langle (\Delta \theta_n-
\langle \Delta \theta_n\rangle )^2\rangle .
\end{eqnarray}
It is expected that by assuming the mixing property of $\alpha_n (\Delta 
\theta_n )$, the variance obeys the diffusion law
\begin{eqnarray}
\sigma_n^2 = 2Dn
\end{eqnarray}
for large $n$, where the diffusion coefficient is given by
$
D=(1/2)\sum_{n=-\infty }^{\infty }C_{|n|}
$
with $C_n=\langle \delta \alpha_n (\Delta \theta_n )\delta \alpha_0 
(\Delta \theta_0)  \rangle,\ (\delta \alpha_n = \alpha_n 
-\langle \alpha \rangle )$.  In the present paper, the diffusion coefficients 
were calculated with the asymptotic form (3.23) for various values of 
$K$ by keeping $\Delta \omega =0.08$.  Results are shown in Fig.~6.  
One observes an anomalous 
enhancement of the diffusion coefficient near $K_*$, the lower 
synchronization-desynchronization transition point.  Figure 7 
displays the relation between the rotation number and the 
diffusion coefficient.  One observes a linear relation 
$D\sim \Delta \Omega $ for $K$ slightly below $K_c$.  This fact 
agrees with the observation in the coupled R\"{o}ssler 
system~\cite{FYK05}.  The numerical result shows $D/\Delta \Omega 
\approx 2.7$. 
\begin{figure}[htb]
\parbox{\halftext}{
    \includegraphics[angle=-90, width=60mm]{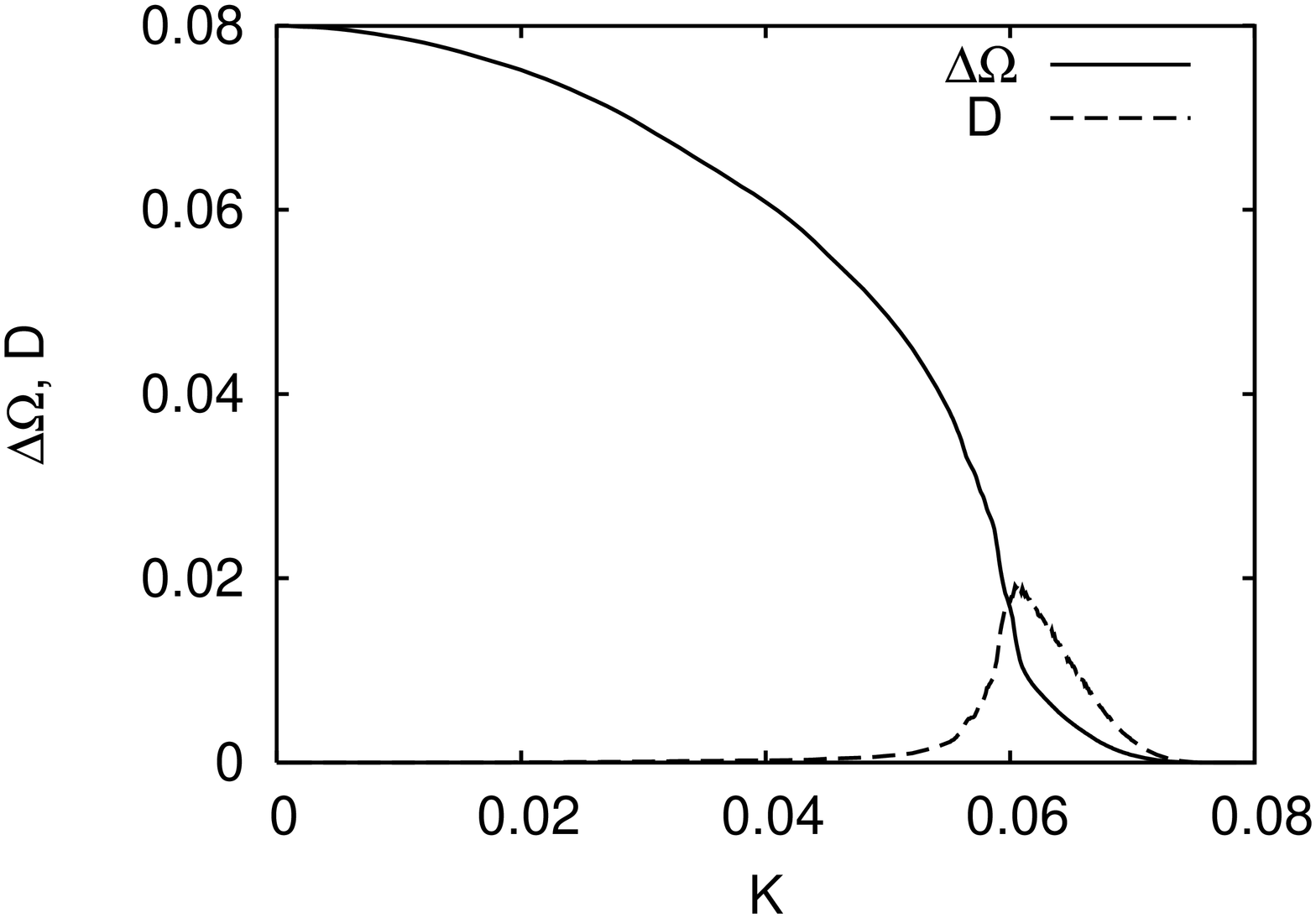}
    \caption{
    The rotation number $\Delta \Omega $ and the phase diffusion 
    coefficient $D$ below the breakdown point $K_c$ of chaotic phase 
    synchronization.  The diffusion coefficient shows an 
    enhancement slightly below the breakdown point of 
    the phase synchronization. 
  }
  \label{fig:6}}
\hfill
\parbox{\halftext}{
  \includegraphics[angle=-90, width=60mm]{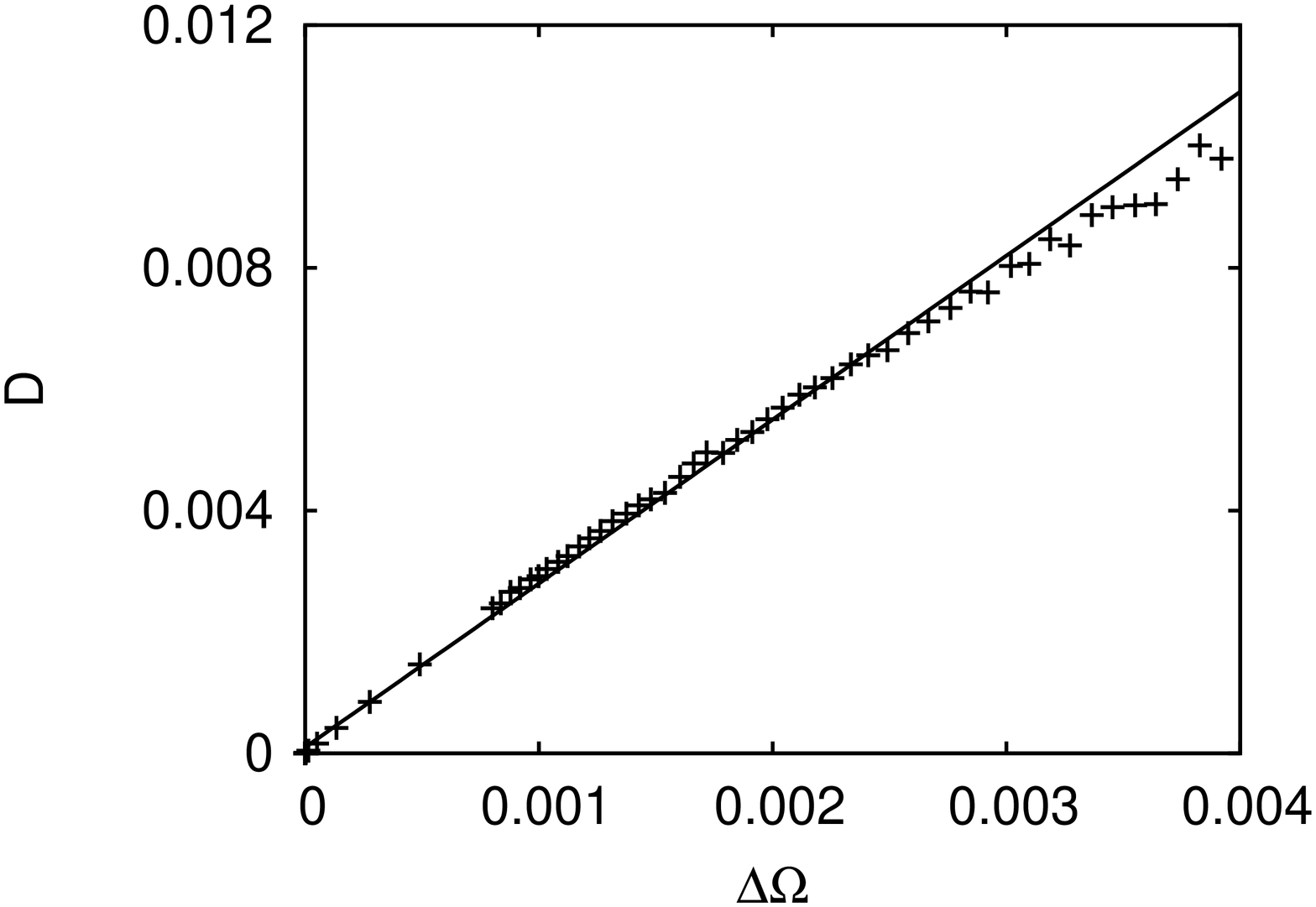}
  \caption{
    The relation between the phase diffusion coefficient and the 
    rotation number near the breakdown point of the chaotic 
    phase synchronization.  One observes a linear 
    relation $D\sim \Delta \Omega $.
  }
  \label{fig:7}}
\end{figure}
\section{Summary and concluding remarks}
In the present paper, we proposed a mapping model of coupled chaotic 
oscillator system composed of element dynamics each of which has 
a periodic oscillation characteristic in an uncoupled case.  
An oscillation has two degrees of freedom, phase and 
amplitude.  Parameters contained in dynamical systems 
described by differential equations of motion for 
describing oscillatory behaviors generically control both the phase and 
the amplitude simultaneously.  On the other hand, the present mapping 
model, the parameter 
$\omega $ mainly controls the phase dynamics and the parameter $a$ 
the amplitude dynamics.  Therefore, in order to study the characteristic 
dynamics 
of the phase and the amplitude separately, the present mapping model 
of oscillatory dynamics is considered to be more convenient to analyze chaotic 
phase synchronization than the 
differential equation system such as the R\"{o}ssler model.
\par
By making use of concrete coupled two maps system, it was shown that the 
present coupled map system shows the phase synchronization-desynchronization 
phenomenon.  We found that when the amplitude synchronization is achieved, 
after the breakdown of the phase synchronization the rotation number turns 
out to show a normal scaling ($3\cdot 20$).  However, since this characteristic is 
quite particular to the element map used in the present paper, in general no 
amplitude synchronization is expected to be observed, and it may be concluded 
that sufficiently near the transition point the anomalous scaling ($3\cdot 21$) 
is ubiquitously observed.  In this sense the present two maps system belongs 
to the universality class of that reported in Ref.~13).  
Furthermore, we studied the phase diffusion coefficient, and observed that 
its critical behaviors are the same as those found in Ref.~14).
\par
So far, mapping models of coupled chaos systems, particularly coupled 
one-dimensional mapping models, have been extensively studied 
to clarify their varieties of dynamical phenomena as well as mathematical 
structures of their dynamical behaviors.  The proposed coupled map model 
composed of chaotic oscillator elements which show well-defined limit-cycle 
type oscillation characteristics is expected to contribute to the 
progress in the study of the phase synchronization-desynchronization 
phenomenon in a wide class of coupled chaos systems.
\section*{Acknowledgments}
The authors thank the members of the Nonequilibrium Dynamics group 
in Graduate School of Informatics at Kyoto University.  
This study was partially supported by the 21st Century COE Program 
``Center Of Excellence for Research and Education on Complex 
Functional Mechanical Systems'' at Kyoto University.


\end{document}